\documentclass[a4,journal]{IEEEtran}
\usepackage[ruled]{algorithm2e}
\usepackage{cite}
\usepackage{graphics}
\usepackage{array}
\usepackage{color}
\usepackage{amsfonts}
\usepackage{import}
\usepackage{tikz}
\usepackage{tabularx}
\usepackage[colorlinks = true,
            linkcolor = blue,
            urlcolor  = blue,
            citecolor = red,
            anchorcolor = blue]{hyperref}
\usetikzlibrary{shapes,arrows}
\usepackage{amsthm,amsmath,amssymb,amsfonts}

\usepackage{xpatch}
\makeatletter
\xpatchcmd{\@thm}{\thm@headpunct{.}}{\thm@headpunct{}}{}{}
\makeatother
\usepackage{multirow}
\newcolumntype{L}{>{\centering\arraybackslash}m{0.7cm}}
\graphicspath{{figures/}}
\usepackage{graphicx, pifont} 

\DeclareMathOperator{\st}{s.t.}

\usetikzlibrary{arrows.meta}
\usetikzlibrary{positioning,fit,backgrounds}
  
\tikzset{%
  >={Latex[width=2mm,length=2mm]},
            base/.style = {rectangle, rounded corners, draw=black,
                           minimum width=1.5cm, minimum height=0.7cm,
                           text centered, font=\sffamily\small},
  activityStarts/.style = {base, fill=black!30},
       startstop/.style = {base, fill=cyan!10},
    activityRuns/.style = {ellipse, minimum width=1cm, minimum height=0.7cm, text centered, font=\sffamily\small, fill=red!25},
         process/.style = {base, minimum width=3cm, fill=yellow!30,
                           font=\sffamily\small},
            a/.style = {rectangle, fill = cyan!10
                           minimum width=2cm, minimum height=0.5cm,
                           text centered, font=\sffamily\small},
}

\begin{document}

\title{A Convergence Predictor Model for Consensus-based Decentralised Energy Markets}
\author{\IEEEauthorblockN{Parikshit Pareek$^1$, L. P. Mohasha Isuru Sampath$^2$, Hung D. Nguyen$^3$, and  Eddy Y. S. Foo$^3$}\vspace{-6mm}
\thanks{$^1$ Theoretical Division (T-5), Los Alamos National Laboratory, NM, USA. Email: pareek@lanl.gov.}
\thanks{$^2$ Institute of High Performance Computing (IHPC), Agency for Science, Technology and Research (A*STAR), Singapore. Email: mohasha\_sampath@ihpc.a-star.edu.sg}
\thanks{$^3$ School of Electrical and Electronic Engineering, NTU, Singapore. Email: \{hunghtd; eddyfoo\}@ntu.edu.sg.}
\thanks{\textcolor{black}{This work was supported in part by the intra-CREATE seed collaboration Fund under Grant NRF2021-ITS008-0017; in part by the National Research Foundation, Singapore.}}}

\maketitle

\begin{abstract}
This letter introduces a convergence prediction model (CPM) for decentralized market clearing mechanisms. The CPM serves as a tool to detect potential cyber-attacks that affect the convergence of the consensus mechanism during ongoing market clearing operations. In this study, we propose a successively elongating Bayesian logistic regression approach to model the probability of convergence of real-time market mechanisms. The CPM utilizes net-power balance among all the prosumers/market participants as a feature for convergence prediction, enabling a low-dimensional model to operate efficiently for all the prosumers concurrently. The results highlight that the proposed CPM has achieved a net false rate of less than 0.01\% for a stressed dataset.
\end{abstract}%

 
\section{Introduction}
Peer-to-peer (P2P) energy trading is a decentralized market framework that allows prosumers to buy and sell electricity directly with one another, without the need for a centralized intermediary such as a utility. A decentralized market-clearing mechanism is required for determining the market price and energy quantities that will be traded in a P2P energy market, while satisfying power balance. Continuous double-auction (CDA) is one of the widely-used market-clearing mechanisms in which the market-clearing entity will determine the market clearing price at which the supply offer order intersects with the demand bid order and matches the total energy supply and demand. CDA is a simple, fair, and efficient process that \textit{guarantees} market-clearing for decentralized markets. 
Recently, CDA has been modified with an iterative scheme in~\cite{JKang_7935397} to achieve consensus between a cluster of electric vehicles and the local aggregator. Further, \emph{iterative} market clearing mechanisms, such as game theory and ADMM, are commonly used in energy markets. In these mechanisms, both sellers and buyers can adjust the energy price and quantities over the iterations. Prosumers solve a local optimization problem based on social-welfare objectives and decide on the most profitable energy exchange or transaction quantities, taking into account the price they receive and the cost of energy produced. Each prosumer solves a constrained optimization problem individually at each iteration, which can be formulated as follows \cite{Mohasha_9530555}:
\begin{subequations}\label{p2p_opt}
\vspace{-0.2em}
\begin{align}
\max_{\boldsymbol{e},\boldsymbol{u}{},\boldsymbol{x}{}} \ & f^{\rm }{} (\boldsymbol{u}{},\boldsymbol{x}{}) +\boldsymbol{\lambda}^\top \boldsymbol{e} \label{p2p_opt_1} \\
\st \ 
& g(\boldsymbol{u}{},\boldsymbol{x}{}) \leq 0 \label{p2p_opt_5} \\
& \underline{p} \leq \mathbf{1}^\top\boldsymbol{e} \leq \overline{p} \label{p2p_opt_3} \\
& \mathbf{1}^\top\boldsymbol{e} =A^{\top}{\rm }\boldsymbol{u} \label{p2p_opt_4} 
\end{align}
\end{subequations}
where, $\boldsymbol{u}{}$ and $\boldsymbol{x}{}$ are the control and state vectors of the prosumer optimization problem, with dimension dependent on the assets possessed. $\boldsymbol{e}$ and $\boldsymbol{\lambda}$ represent P2P energy transaction quantities and price agreements with other prosumers in the decentralized market. The objective function \eqref{p2p_opt_1} represents the \textit{maximization} of overall welfare and profit from energy sales, subject to the constraints of the prosumer assets \eqref{p2p_opt_5} and the market \eqref{p2p_opt_3}. The power balance is represented in \eqref{p2p_opt_4}, where matrix $A$ maps the power production and consumption of assets to energy exchanges. Here, $\boldsymbol{e}$ is determined while solving \eqref{p2p_opt} for a given $\boldsymbol{\lambda}$. Then, $\boldsymbol{\lambda}$ is updated typically using a linear rule such that the negotiations between trading pairs are converging towards an \textit{equilibrium}. The \textit{consensus protocol} which governs the negotiation process must be strictly followed while information exchange between market participants to guarantee the convergence. Then, at the termination, the efficiency, security and reliability of the energy trading process, such as optimal allocation of resources~\cite{Mohasha_8913591}, satisfying power/energy balance~\cite{Mohasha_8913591,Mohasha_9530555}, respecting the engineering constraints of the grid and assets~\cite{Mohasha_9530555}, etc., are fulfilled. 

As such, accurate information exchange between market participants is crucial for the convergence/termination of the market clearing process. False data shared by one or more agents that violate the consensus protocol hinders its ability to establish an agreement on resource allocation, which can be discouraging for all the market participants. Data validation and verification mechanisms can ensure the accuracy and integrity of the data being exchanged. Penalizing agents for intentionally sending false data can deter such behavior and promote market integrity. Consequently, it is beneficial if the market clearing entity possesses sophisticated methods to determine the likelihood of market clearing based on the shared data by participating agents over the iterations. This study aims to devise a tool that provides a probabilistic assurance of market clearing convergence.
The main contributions of this letter can be summarized as:

\begin{itemize}
    \item Formulating market convergence prediction problem in probabilistic domain using Bayesian Logistic Regression (BLR). Using net power-balance as the feature for convergence prediction allows low-dimensional model to work for all prosumers simultaneously. 
    \item Proposing a successively elongating BLR model for predicting convergence of an ongoing P2P market mechanism. The successively increasing length of input vector allows to observe change in convergence probability during an ongoing market clearance. 
\end{itemize}

\color{black}

\section{Convergence Predictor Model}
This section presents the proposed CPM, the BLR-based model, for predicting the convergence of the energy trading mechanism, along with the approach used to understand time-series behavior. 


In BLR, we model event occurrence probability using log-odds as a linear combination of independent variables. The parameters of the model are estimated using Bayesian techniques. This allows us to make probabilistic predictions and incorporate prior knowledge. In this work, BLR is preferred over other models due to its ability to regularize the model and prevent over-fitting. Also, BLR's probabilistic predictions can be useful in the decision-making task as the confidence level is important for a CPM. Moreover, decentralized energy markets such as P2P markets works with sharing of two data sets: i) individual participant bids and ii) convergence parameters~\cite{Mohasha_9530555}. These data provide comprehensive information on the market mechanism, making them suitable to be used as input for a CPM. However, utilizing the entire feature space poses several challenges, such as difficulties in securely storing historical data, verifying the accuracy of individual prosumer bids, and issues related to the ``\textit{curse-of-dimensionality}." Therefore, we propose to use the power balance gap\footnote{Any convergence indicator can be used as a feature in a general decentralized market mechanism.} \eqref{p2p_opt_4} as the feature for our CPM model, reducing data storage and feature space dimensions. The convergence criteria \eqref{p2p_opt} considers the impact of all prosumer data, by summing their power-bids, thus is suitable for dimensionality reduction in the proposed CPM.

Let, $x\in \mathbb{R}^{D}$ be a vectors with $D$ input features\footnote{$x$ is vector of power-balance gaps after $D$ iterations of market mechanism.}, $\boldsymbol \alpha \in \mathbb{R}^{D}$ be the model weight vector, and $b$ be the model bias term. The outcome probability, given the features and model hyperparameters, is given by $\mathbb{P}(y | x, \boldsymbol \alpha, b) = \frac{1}{1 + \exp(-y(\boldsymbol \alpha^\top x + b))}$ where $y=1$ for converged and $y=0$ for non-converged cases. The term $\boldsymbol \alpha^\top x + b$ is known as the logistic function, which maps input to the range $[0,1]$. In the Bayesian paradigm, both the weight vector and bias are considered random variables with prior distributions $\mathbb{P}(\boldsymbol \alpha)$ and $\mathbb{P}(b)$, respectively, to incorporate prior understanding of the model. Using Baye's rule and input data $X \in \mathbb{R}^{N\times D}$, the posterior distributions of the weight vector and bias can be given as:
\begin{align}\label{eq:prob_wb}
\small
\mathbb{P}(\boldsymbol \alpha | y, X) = \frac{\mathbb{P}(y | X, \boldsymbol \alpha) \mathbb{P}(\boldsymbol \alpha)}{\mathbb{P}(y | X)}; \ \
\mathbb{P}(b | y, X) = \frac{\mathbb{P}(y | X, b) \mathbb{P}(b)}{\mathbb{P}(y | X)}
\end{align}

The right-hand side of the equation provides the probability of model hyperparameters given the input data and outcome (known as learning/training data). The marginal likelihood, $\mathbb{P}(y | X)$, is obtained by integrating the joint likelihood $\mathbb{P}(y, \boldsymbol \alpha, b | X)$ over all possible values of $\boldsymbol \alpha$ and $b$. However, this integration is an intractable problem due to the high-dimensional space involved ($D$-dimensional) and the absence of closed-form \cite{schmidt2020bayesian,makalic2015simple}. A solution to this challenge is the use of a "horseshoe estimator" which assigns "horseshoe" prior to the model hyperparameters $\mathbb{P}(\boldsymbol \alpha)$ and $\mathbb{P}(b)$. These priors encourage most parameters to be close to zero while allowing flexibility for the model to fit the data. We use the {\fontfamily{lmtt}\selectfont bayesreg} package for the building CPM and implementation details can be found in \cite{makalic2016high}. Using \eqref{eq:prob_wb}, we can obtain the posterior distributions. The probability of a market converging to $y=1$ is given by $\mathbb{P}(y_{\text{new}}=1 | \mathbf{x}_{\text{new}}, \boldsymbol \alpha\star, b_\star) = {1}/\{1 + \exp(-(\boldsymbol \alpha_\star^\top \mathbf{x}_{\text {new}} + b_\star))\}$. The probability of failure (non-converging) can be obtained as $1-\mathbb{P}(y_{\text{new}}=1 | \mathbf{x}_{\text{new}}, \boldsymbol \alpha_\star, b_\star)$. Thus, BLR provides a probabilistic framework for answering the question of \textit{what is the probability that a given market mechanism will converge?}

\begin{figure}[t]
 \centering
    \resizebox{0.8\columnwidth}{3.5cm}{
    
\tikzset{every picture/.style={line width=0.75pt}} 
\begin{tikzpicture}[x=0.75pt,y=0.75pt,yscale=-1,xscale=1]

\draw  [fill={rgb, 255:red, 117; green, 117; blue, 117 }  ,fill opacity=0.6 ] (100,135) .. controls (100,135) and (100,135) .. (100,135) -- (397.36,135) .. controls (397.36,135) and (397.36,135) .. (397.36,135) -- (397.36,175) .. controls (397.36,175) and (397.36,175) .. (397.36,175) -- (100,175) .. controls (100,175) and (100,175) .. (100,175) -- cycle ;
\draw  [fill={rgb, 255:red, 180; green, 190; blue, 40 }  ,fill opacity=0.6 ] (100,135) .. controls (100,135) and (100,135) .. (100,135) -- (199.12,135) .. controls (199.12,135) and (199.12,135) .. (199.12,135) -- (199.12,175) .. controls (199.12,175) and (199.12,175) .. (199.12,175) -- (100,175) .. controls (100,175) and (100,175) .. (100,175) -- cycle ;
\draw  [fill={rgb, 255:red, 210; green, 240; blue, 40 }  ,fill opacity=0.6 ] (99.77,135.28) .. controls (99.77,135.28) and (99.77,135.28) .. (99.77,135.28) -- (149.34,134.71) .. controls (149.34,134.71) and (149.34,134.71) .. (149.34,134.71) -- (149.78,174.72) .. controls (149.78,174.72) and (149.78,174.72) .. (149.78,174.72) -- (100.22,175.29) .. controls (100.22,175.29) and (100.22,175.29) .. (100.22,175.29) -- cycle ;
\draw    (100,175) -- (425.1,175)[color={rgb, 255:red, 0; green, 0; blue, 0 }  ] ;
\draw [shift={(427.1,175)}, rotate = 180] [color={rgb, 255:red, 0; green, 0; blue, 0 }  ][line width=0.75]    (10.93,-3.29) .. controls (6.95,-1.4) and (3.31,-0.3) .. (0,0) .. controls (3.31,0.3) and (6.95,1.4) .. (10.93,3.29)   ;
\draw [line width=1.5]  [dash pattern={on 3.75pt off 3pt on 7.5pt off 1.5pt}]  (149.73,173.16) -- (149.56,200) ;
\draw [line width=1.5]  [dash pattern={on 3.75pt off 3pt on 7.5pt off 1.5pt}]  (199.12,170) -- (199.12,225) ;
\draw [color={rgb, 255:red, 14; green, 25; blue, 235 }  ,draw opacity=1 ]   (102,200) -- (147.56,200) ;
\draw [shift={(149.56,200)}, rotate = 180] [color={rgb, 255:red, 14; green, 25; blue, 235 }  ,draw opacity=1 ][line width=0.75]    (10.93,-3.29) .. controls (6.95,-1.4) and (3.31,-0.3) .. (0,0) .. controls (3.31,0.3) and (6.95,1.4) .. (10.93,3.29)   ;
\draw [shift={(100,200)}, rotate = 0] [color={rgb, 255:red, 14; green, 25; blue, 235 }  ,draw opacity=1 ][line width=0.75]    (10.93,-3.29) .. controls (6.95,-1.4) and (3.31,-0.3) .. (0,0) .. controls (3.31,0.3) and (6.95,1.4) .. (10.93,3.29)   ;
\draw [color={rgb, 255:red, 14; green, 25; blue, 235 }  ,draw opacity=1 ]   (104,225) -- (195,225) ;
\draw [shift={(197,225)}, rotate = 180] [color={rgb, 255:red, 14; green, 25; blue, 235 }  ,draw opacity=1 ][line width=0.75]    (10.93,-3.29) .. controls (6.95,-1.4) and (3.31,-0.3) .. (0,0) .. controls (3.31,0.3) and (6.95,1.4) .. (10.93,3.29)   ;
\draw [shift={(102,225)}, rotate = 0] [color={rgb, 255:red, 14; green, 25; blue, 235 }  ,draw opacity=1 ][line width=0.75]    (10.93,-3.29) .. controls (6.95,-1.4) and (3.31,-0.3) .. (0,0) .. controls (3.31,0.3) and (6.95,1.4) .. (10.93,3.29)   ;
\draw [line width=1.5]  [dash pattern={on 3.75pt off 3pt on 7.5pt off 1.5pt}]  (397.53,173.16) -- (397.36,250) ;
\draw [line width=1.5]  [dash pattern={on 3.75pt off 3pt on 7.5pt off 1.5pt}]  (100.22,175.28) -- (100.06,252.12) ;
\draw [color={rgb, 255:red, 13; green, 15; blue, 235 }  ,draw opacity=1 ] [dash pattern={on 4.5pt off 4.5pt}]  (102,145) -- (147.56,145) ;
\draw [shift={(149.56,145)}, rotate = 180] [color={rgb, 255:red, 13; green, 15; blue, 235 }  ,draw opacity=1 ][line width=0.75]    (10.93,-3.29) .. controls (6.95,-1.4) and (3.31,-0.3) .. (0,0) .. controls (3.31,0.3) and (6.95,1.4) .. (10.93,3.29)   ;
\draw [shift={(100,145)}, rotate = 0] [color={rgb, 255:red, 13; green, 15; blue, 235 }  ,draw opacity=1 ][line width=0.75]    (10.93,-3.29) .. controls (6.95,-1.4) and (3.31,-0.3) .. (0,0) .. controls (3.31,0.3) and (6.95,1.4) .. (10.93,3.29)   ;
\draw    (377.54,267) -- (397.36,267)[color={rgb, 255:red, 235; green, 0; blue, 0 }  ] ;
\draw [shift={(397.36,267)}, rotate = 180] [color={rgb, 255:red, 235; green, 0; blue, 0 }  ][line width=0.75]    (0,5.59) -- (0,-5.59)(10.93,-3.29) .. controls (6.95,-1.4) and (3.31,-0.3) .. (0,0) .. controls (3.31,0.3) and (6.95,1.4) .. (10.93,3.29)   ;
\draw    (119.82,267) -- (100,267)[color={rgb, 255:red, 235; green, 0; blue, 0 }  ] ;
\draw [shift={(100,267)}, rotate = 360] [color={rgb, 255:red, 235; green, 0; blue, 0 }  ][line width=0.75]    (0,5.59) -- (0,-5.59)(10.93,-3.29) .. controls (6.95,-1.4) and (3.31,-0.3) .. (0,0) .. controls (3.31,0.3) and (6.95,1.4) .. (10.93,3.29)   ;

\draw [color={rgb, 255:red, 14; green, 25; blue, 235 }  ,draw opacity=1 ]   (102,250) -- (393,250) ;
\draw [shift={(395,250)}, rotate = 180] [color={rgb, 255:red, 14; green, 25; blue, 235 }  ,draw opacity=1 ][line width=0.75]    (10.93,-3.29) .. controls (6.95,-1.4) and (3.31,-0.3) .. (0,0) .. controls (3.31,0.3) and (6.95,1.4) .. (10.93,3.29)   ;
\draw [shift={(100,250)}, rotate = 0] [color={rgb, 255:red, 14; green, 25; blue, 235 }  ,draw opacity=1 ][line width=0.75]    (10.93,-3.29) .. controls (6.95,-1.4) and (3.31,-0.3) .. (0,0) .. controls (3.31,0.3) and (6.95,1.4) .. (10.93,3.29)   ;
\draw [color={rgb, 255:red, 17; green, 148; blue, 35 }  ,draw opacity=1 ]   (149.34,134.71) -- (149.34,109.71) ;
\draw [shift={(149.34,109.71)}, rotate = 270] [color={rgb, 255:red, 17; green, 148; blue, 35 }  ,draw opacity=1 ][fill={rgb, 255:red, 17; green, 148; blue, 35 }  ,fill opacity=1 ][line width=0.75]      (0, 0) circle [x radius= 3.35, y radius= 3.35]   ;
\draw [color={rgb, 255:red, 17; green, 148; blue, 35 }  ,draw opacity=1 ]   (199.12,135) -- (199.12,110) ;
\draw [shift={(199.12,110)}, rotate = 270] [color={rgb, 255:red, 17; green, 148; blue, 35 }  ,draw opacity=1 ][fill={rgb, 255:red, 17; green, 148; blue, 35 }  ,fill opacity=1 ][line width=0.75]      (0, 0) circle [x radius= 3.35, y radius= 3.35]   ;
\draw [color={rgb, 255:red, 17; green, 148; blue, 35 }  ,draw opacity=1 ]   (397.36,135) -- (397.36,110) ;
\draw [shift={(397.36,110)}, rotate = 270] [color={rgb, 255:red, 17; green, 148; blue, 35 }  ,draw opacity=1 ][fill={rgb, 255:red, 17; green, 148; blue, 35 }  ,fill opacity=1 ][line width=0.75]      (0, 0) circle [x radius= 3.35, y radius= 3.35]   ;
\draw (119.69,258) node [anchor=north west][inner sep=0.75pt]  [font=\small,color={rgb, 255:red, 231; green, 25; blue, 12 }  ,opacity=1 ]  {$ \begin{array}{l}
\mathrm{Length\ of\ feature\ vector\ } \mathbf{x} \ \mathrm{for\ } M^{\rm th}\ \mathrm{model}\\
\end{array}$};
\draw (118,185) node [anchor=north west][inner sep=0.75pt]  [color={rgb, 255:red, 14; green, 25; blue, 235 }  ,opacity=1 ]  {$w_{1}$};
\draw (110.79,147.4) node [anchor=north west][inner sep=0.75pt]  [color={rgb, 255:red, 0; green, 9; blue, 255 }  ,opacity=1 ]  {$\Delta w$};
\draw (161,210) node [anchor=north west][inner sep=0.75pt]  [color={rgb, 255:red, 14; green, 25; blue, 235 }  ,opacity=1 ]  {$w_{2}$};
\draw (248.2,226.4) node [anchor=north west][inner sep=0.75pt]  [color={rgb, 255:red, 14; green, 25; blue, 235 }  ,opacity=1 ]  {$w^{}_{M} =\ M\Delta w$};
\draw (403.2,183.14) node [anchor=north west][inner sep=0.75pt]  [rotate=-0.32]  {$\#\ Iter.$};
\draw (254.2,197.4) node [anchor=north west][inner sep=0.75pt]  [color={rgb, 255:red, 231; green, 25; blue, 12 }  ,opacity=1 ]  {$D=\ M\Delta w$};
\draw (146,86.4) node [anchor=north west][inner sep=0.75pt]  [color={rgb, 255:red, 14; green, 25; blue, 235 }  ,opacity=1 ]  {$\mathbb{P}_{1}$};
\draw (195,86.4) node [anchor=north west][inner sep=0.75pt]  [color={rgb, 255:red, 14; green, 25; blue, 235 }  ,opacity=1 ]  {$\mathbb{P}_{2}$};
\draw (393,86.4) node [anchor=north west][inner sep=0.75pt]  [color={rgb, 255:red, 14; green, 25; blue, 235 }  ,opacity=1 ]  {$\mathbb{P}_{M}$};
\draw (216,115.4) node [anchor=north west][inner sep=0.75pt]  [color={rgb, 255:red, 231; green, 25; blue, 12 }, font=\small]  {$\mathbb{P}_{m\ } =\mathbb{P}\left( y=1|\mathbf{x} \ \in \mathbb{R}^{m\Delta w}\right) \ $};
\end{tikzpicture}
}
\vspace{-2mm}
\caption{Idea of successively elongating BLR model for the proposed CPM.}
\vspace{-1.5em}
\label{fig:model}
\end{figure}
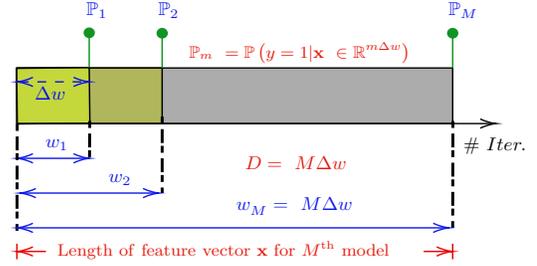

Fig.~\ref{fig:model} shows the successively elongating BLR model approach to capture time-series behaviour of the ongoing market mechanism. In this approach, $n$-th model will take input vector of length $n\Delta w$. In this manner, the variations in convergence probability can be analysed as a time series. Total number of such models depends upon $\Delta w$, which is selected using sensitivity analysis over historical market dataset.


\color{black}

\section{Results and Discussion}
In this study, we evaluate the performance of the proposed CPM for a P2P energy market mechanism using False Negative Rate (FNR), False Positive Rate (FPR), and Matthew's Correlation Coefficient (MCC) \cite{bishop2006pattern}. FNR measures the proportion of actual negatives misclassified as positives and FPR measures the proportion of actual positives misclassified as negatives. MCC, a measure of correlation between predicted and actual binary classifications, ranges from -1 to 1, where 1 represents a perfect prediction, -1 represents total disagreement and 0 represents no correlation. A good performance of the model is indicated by a combination of a low FNR, a low FPR, and a high MCC. In the proposed CPM, FPR represents the ratio of instances of cyber-attacks or malfunctions that went unchecked, while FNR represents the ratio of premature terminations of a secure market clearing mechanism. Moreover, the testing uses two different datasets obtained considering the decentralized market clearing process in \eqref{p2p_opt}. In Data\#1 any of the three-prosumer is considered being under attack randomly from the beginning of the market process. However, in Data\#2 attack is considered to be starting randomly between iteration number between 15 and 55 and to a random prosumer (1-3). Therefore, the Data\#2 should be considered a stressed dataset.

\begin{figure}[t]
    \centering
    \includegraphics[width=\columnwidth]{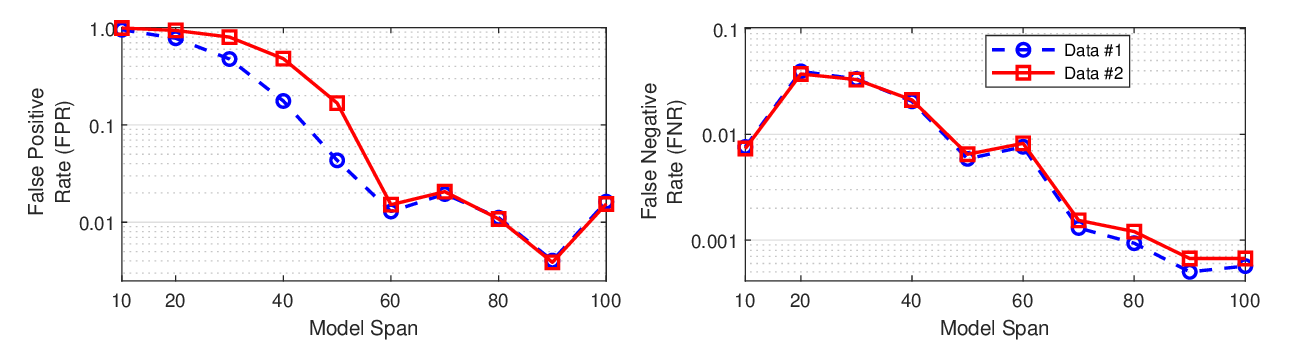}
    \vspace{-7.0mm}
	\caption{FPR and FNR for normal and stressed data sets of market clearing with 3 prosumers, with different model spans. The results indicate lowest false rates with model span of 90 iterations.}
    \label{fig:FPR_FNR}
         \vspace{-1em}
\end{figure}

\begin{figure}[t]
    \centering
    \includegraphics[width=\columnwidth]{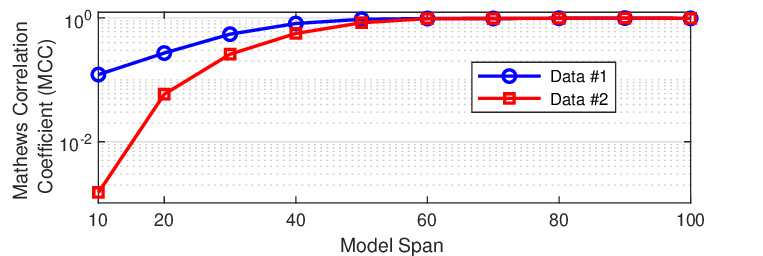}
    \vspace{-8.0mm}
	\caption{Performance quantification of the proposed CPM using MCC for different datasets at different model spans.}
    \label{fig:MCC}
    \vspace{-1.3em}
\end{figure}

Fig. \ref{fig:FPR_FNR} shows the FPR and FNR obtained using the proposed CPM with different window sizes for both normal (Data\#1) and stressed (Data\#2) dataset. It is observed that the prediction accuracy of the normal dataset is higher compared to the stressed one as it gives a lower FPR. However, with larger model span ($\geq$ 60) FPR  reaches lower than 0.01 with minimum being 0.0038 at the model span of 90 for both the datasets. Similarly, the FNR values also lowest at the model span of 90 iterations. This shows that proposed CPM has been been able to achieve acceptably lower false prediction rates (both positive and negative). Now, we measure the performance of the proposed classification model using MCC with different model spans. As illustrated in Fig.~\ref{fig:MCC}, it is clear that MCC is lower for stressed data set (Data \#2) at smaller model spans. However, for both data sets, the MCC goes near unity for model spans greater than 60. This higher value of MCC in Fig.~\ref{fig:MCC}, with results in Fig.~\ref{fig:FPR_FNR}, indicates highly accurate performance of the proposed CPM for identification of compromised market mechanisms.

\begin{figure}[t]
\centering
\includegraphics[width=\columnwidth]{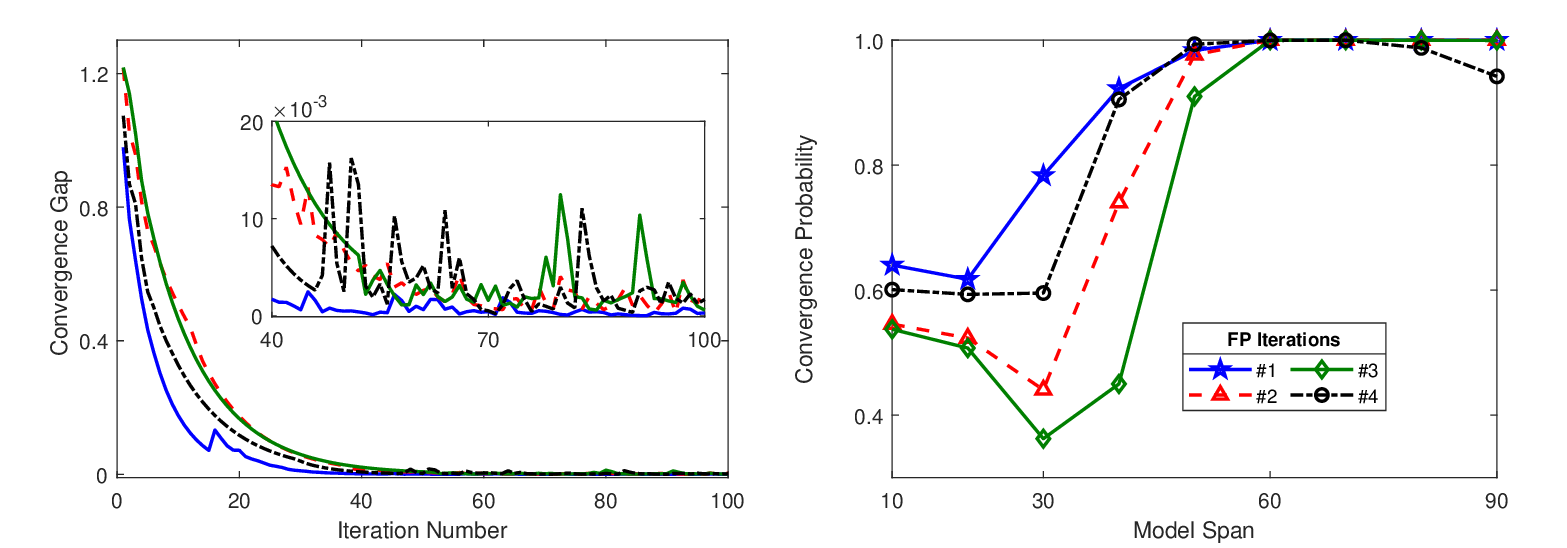}
\vspace{-7.0mm}
\caption{False positive analysis for a few samples with minimum convergence gap as ${1.77;26.0;58.90;22.70}\times 10^{-5}$ for \#1-\#4 respectively.}
\label{fig:false_positive}
\vspace{-1em}
\end{figure}

Although, the proposed CPM has low failure rate, yet it is important to analyse failure instances. For this, we plot four randomly selected false positive instances, that was labeled safe wrongly by the CPM, in Fig.~\ref{fig:false_positive}. Firstly, all these instances show smooth convergence trajectories (convergence gap variation against iteration number), with a few abrupt shifts as illustrated on the left sub-figure of Fig.~\ref{fig:false_positive}. Further, in inset, it is clearly visible that all these four instances show variability in convergence gap of the order of $10^{-3}$. Furthermore, the minimum value of power balance (convergence tolerance) obtained in these is in the order of $10^{-5}$. These properties (smoothness, lower variability, and small minimum value) of failed data samples imply that the proposed CPM fails when market mechanism under attack behaves closely to the properties of a safe and converging market clearing process\footnote{Notable, if an attack stays within (or close to) the convergence tolerance of the P2P market clearing mechanism, the system failure risk is low. Because convergence gap is the difference in power supply and demand balance, and the system operator has load matching reserves for small load differences, such as $10^{-5}$ or even $10^{-3}$ p.u.}. Lastly, it is important to conduct a sensitivity study to observe the proposed CPM performance with different window sizes $\Delta W$. Table~\ref{tab:span} reports the effect of $\Delta W$ on MCC values for Data\#2. It is evident that the proposed CPM shows similar performance with all different model spans. This indicates that the idea of proposed \textit{window starching approach} has the potential to make the proposed CPM span agnostic.

\begin{table}[t]
\centering
\caption{Effect of $\Delta W$ on MCC Values with Testing on Data \#2} 
\vspace{-0.5em}
\begin{tabular}{c|ccccc}
     & \multicolumn{5}{c}{Model Span: $M\Delta w$}        \\ \cline{2-6} 
$\Delta w$ & 20    & 40    & 60    & 80    & 100    \\
\hline
5    & 0.064 & 0.562 & 0.974 & 0.986 & 0.974 \\
10    & 0.059 & 0.561 & 0.977 & 0.988 & 0.984 \\
20     & 0.066 & 0.562 & 0.969 & 0.984 & 0.991\\
\hline
\end{tabular}
\vspace{-1.5em}
\label{tab:span}
\end{table}

\section{Conclusions}
This letter presents a convergence prediction model (CPM) designed for decentralized market clearing mechanisms. The CPM effectively detects potential cyber-attacks during ongoing market operations, enhancing the security of the system. By employing a successively elongating Bayesian logistic regression (BLR) approach, we accurately model the probability of convergence in real-time market mechanisms. The validation study considers two types of datasets comprising different cyber-attack considerations. The prediction accuracy of the proposed CPM was verified to be acceptable based on three standard metrics: false positive rate, false negative rate, and Matthew's correlation coefficient. Future work will investigate additional measures, such as incorporating anomaly detection techniques, to identify sophisticated cyber-attacks.

\bibliographystyle{IEEEtran}
\bibliography{main}

\end{document}